\documentclass[11pt]{article}

\usepackage{amsthm,amsmath,amssymb}
\usepackage{natbib}
\usepackage{multirow}
\usepackage[pdftex]{graphicx}
\usepackage{subfigure}
\usepackage{makecell}
\usepackage{booktabs}
\usepackage{array}
\usepackage{fullpage}
\usepackage{url}
\usepackage{algorithm}
\usepackage{algorithmic}
\usepackage{bm}
\usepackage{smile}
\usepackage{wrapfig}
\usepackage{lipsum}
\usepackage{mathrsfs}
\usepackage{dsfont}
\usepackage{titling}
\usepackage{epstopdf}
\usepackage{setspace}

\usepackage{natbib}
\usepackage{multirow}

\usepackage[usenames,dvipsnames,svgnames,table]{xcolor}
\usepackage[colorlinks=true,
            linkcolor=blue,
            urlcolor=blue,
            citecolor=blue]{hyperref}

\newcommand{\R}{{R}}
\newcommand{\RB}{{RB}}
\newcommand{\CBPS}{{CBPS}}
\newcommand{\hCBPS}{{HD-CBPS}}
\newcommand{\AIPW}{{AIPW}}
\newcommand{\AIPWNR}{{AIPW-NR}}
\newcommand{\DS}{{D-SELECT}}

\begin{document}

\title{\huge Robust Estimation of Causal Effects via High-Dimensional
  Covariate Balancing Propensity Score}

\author{Yang Ning\thanks{Department of Statistical Science, Cornell University, Ithaca, New York 14853, U.S.A. ; e-mail: \texttt{yn265@cornell.edu}.}~~~~~Sida Peng\thanks{Department of Economics, Cornell University, Ithaca, New York 14853, U.S.A. ; e-mail: \texttt{sp947@cornell.edu}.}~~~~ Kosuke Imai\thanks{Department of Government and Department of Statistics, Harvard University, Cambridge, Massachusetts 02138, U.S.A. e-mail: \texttt{imai@harvard.edu}. }
}

\date{}

\maketitle

\vspace{-0.5in}

\begin{abstract}
  In this paper, we propose a robust method to estimate the average
  treatment effects in observational studies when the number of
  potential confounders is possibly much greater than the sample size.
  We first use a class of penalized $M$-estimators for the propensity
  score and outcome models. We then calibrate the initial estimate of
  the propensity score by balancing a carefully selected subset of
  covariates that are predictive of the outcome.  Finally, the
  estimated propensity score is used to construct the inverse
  probability weighting estimator.  We prove that the proposed
  estimator, which has the sample boundedness property, is root-$n$
  consistent, asymptotically normal, and semiparametrically efficient
  when the propensity score model is correctly specified and the
  outcome model is linear in covariates.  More importantly, we show
  that our estimator remains root-$n$ consistent and asymptotically
  normal so long as either the propensity score model or the outcome
  model is correctly specified.  We provide valid confidence
  intervals in both cases and further extend these results to the case
  where the outcome model is a generalized linear model.  In
  simulation studies, we find that the proposed methodology often
  estimates the average treatment effect more accurately than the
  existing methods.  We also present an empirical application, in
  which we estimate the average causal effect of college attendance on
  adulthood political participation.  Open-source software is
  available for implementing the proposed methodology.\footnote{An earlier version of this paper available online at June 2017 has a different title ``High Dimensional Propensity Score Estimation via Covariate Balancing".}
\end{abstract}


\section{Introduction}

Propensity score of \citet{rose:rubi:83} plays a central role in the
estimation of causal effects in observational studies \citep[see
e.g.,][for extensions to a non-binary
treatment]{imbe:00,imai:vand:04}.  In particular, matching and
weighting methods based on propensity score have become part of
applied researchers' standard toolkit across many scientific
disciplines \citep[see e.g.,][]{lunc:davi:04,rubi:06}.  One important
challenge, which is becoming increasingly common as the amount of
available data grows, is the question of how to incorporate a large
number of potential confounders.  For example, \citet{schn:etal:09}
considers a total of several thousand candidate confounders obtained
from the health care claims data.

In this paper, we propose a robust method to estimate the average
treatment effect (ATE) in observational studies when the number of
potential confounders is possibly much greater than the sample
size. In particular, under the standard assumption of strong
ignorability, we propose to estimate the propensity score by balancing
covariates in high-dimensional settings. The proposed method consists
of several steps. We first obtain an initial estimator of the
propensity score model by maximizing a penalized generalized
quasi-likelihood, which depends on a user-specified weight
function. Next, we apply the weighted least squares method to fit the
outcome model.  We show that the two weight functions critically
determine the performance of the proposed estimator under model
misspecification. Third, we refine the initial estimate of the
propensity score by balancing a carefully selected set of observed
covariates that are predictive of the outcome. Finally, the estimated
propensity score is used to construct the inverse probability
weighting estimator of the ATE.

We prove that under mild conditions the proposed estimator of ATE is
root-$n$ consistent, asymptotically normal, and semiparametrically
efficient, when the propensity score model is correctly specified and
the outcome model is linear in covariates. This result holds for a
broad class of weight functions used in the initial estimation of
propensity score.  However, the proposed estimator typically has a
slower rate of convergence under misspecified models.  To address this
problem, we show that by carefully choosing the weight functions the
proposed estimator remains root-$n$ consistent and asymptotically
normal so long as either the propensity score model or the outcome
model is correctly specified.  The proposed estimator has the double
robustness and sample boundedness properties and comes with honest
confidence intervals.  Finally, we extend these theoretical results to
the case where the outcome model is a generalized linear model in
order to allow for nonlinearity.

We emphasize that the proposed methodology does not require the
variable selection consistency of either the propensity score model or
the outcome model.  This is because our goal is to estimate causal
effects rather than the coefficients of propensity score and outcome
models.  The covariate balancing step of our methodology removes the
bias that results from the failure to select some covariates.

The proposed methodology, which we call the high-dimensional covariate
balancing propensity score (\hCBPS), builds on three strands of
research that have recently emerged in the causal inference
literature. In the following, we briefly highlight the differences
between the \hCBPS\ and the existing methods. Section~\ref{seccompare}
further presents a more detailed comparison.

First, a number of researchers have recently proposed to estimate the
ATE by optimizing covariate balance between the treatment and control
groups
\citep[e.g.,][]{hain:12,grah:pint:egel:12,MR3153941,chan:yam:zhan:16,
  zubizarreta2015stable,zhao2016covariate,fan2016}. It has been shown,
both theoretically and empirically, that these approaches can
significantly improve the efficiency and robustness of standard
propensity score methods.  The proposed \hCBPS\ methodology extends
the covariate balancing propensity score (\CBPS) methodology of
\citet{MR3153941} and \citet{fan2016} to the high-dimensional
settings, in which the number of potential confounders is possibly
greater than the sample size. While the original CBPS methodology
estimates the propensity score by balancing all covariates, this is
not an effective strategy in high-dimensional settings because the number of covariates is too large.  To address
this issue, we propose a weak covariate balancing approach, which yields a root-$n$ consistent estimator in high-dimensional settings. 

Second, we contribute to the growing literature on the estimation of
the ATE in high-dimensional settings. \cite{belloni2014inference}
proposed a double selection approach to infer the coefficient of
treatment variable in a partially linear model under the assumption
that both the outcome and treatment models are sparse.
\cite{farrell2015robust}, \cite{belloni2013program}, and
\cite{Chernozhukov2016} extended the augmented inverse probability
weighting (\AIPW) estimator of \cite{robins1994estimation} to
high-dimensional settings. A common characteristic of these methods is
to first estimate the nuisance parameters (e.g., propensity score)
typically by the penalized maximum likelihood and then estimate the
ATE by solving the efficient score function.
Different from this line of work, we rely on the covariate balancing
strategy for estimating the propensity score model and use the
Horvitz-Thompson estimator \citep{horv:thom:52} for inferring the ATE
without employing the AIPW estimator. As elaborated in
Section~\ref{seccompare}, the robustness of the asymptotic
distributions of our estimator to model misspecification is the main
advantage over these existing methods. Most recently,
\cite{tan2017regularized,tan2018model} proposed a penalized calibrated
propensity score method and studied its robustness to model
misspecification. Unlike this method, our approach is based on the covariate balancing.  See
Section~\ref{seccompare} for a detailed comparison.

Finally, \hCBPS\ is related to the recently proposed approximate
residual balancing method \citep{athey2016efficient}, which unlike our
methodology has an advantage of not requiring the formulation of a
propensity score model.  While the approximate residual balancing
method requires the outcome model to be linear in covariates, \hCBPS\
can yield a consistent and asymptotically normal estimator even under
the misspecification of the outcome model so long as the propensity
score is correctly specified. In addition, we show that
the proposed method and its asymptotic theory can be extended to the
case, in which the outcome variable follows a generalized linear
model.  This also overcomes the same limitation of the original \CBPS\
estimator \citep{MR3153941,fan2016}.  Finally, we also argue that the
estimation of the propensity score may help scientists better
understand the treatment assignment mechanism
\citep[e.g.,][]{rubi:08}, when it is correctly specified.  Open-source software package, {\sf CBPS},
is available for implementing our proposed methodology
\citep{fong:etal:18}.

Throughout the paper, we use the following notation. For
$v=(v_1,...,v_d)^\top \in \mathbb{R}^d$, and $1 \leq q \leq \infty$,
we define $\|{v}\|_q=(\sum_{i=1}^d |v_i|^q)^{1/q}$,
$\|{v}\|_0=|\textrm{supp}(v)|$, where
$\textrm{supp}(v)=\{j: v_j\neq 0\}$ and $|A|$ is the cardinality of a
set $A$. Denote $v^{\otimes
  2}=vv^\top$. 
If the matrix $M$ is symmetric, then $\lambda_{\min}(M)$ and
$\lambda_{\max}(M)$ are the minimal and maximal eigenvalues of
$M$. 
For $S\subseteq \{1,...,d\}$, let $v_S=\{v_j: j\in S\}$ and $S^c$ be
the complement of $S$. For two positive sequences $a_n$ and $b_n$, we
write $a_n\asymp b_n$ if $C\leq a_n/b_n\leq C'$ for some
$C,C'>0$. Similarly, we use $a_n\lesssim b_n$ to denote $a_n\leq Cb_n$
for some constant $C>0$. A random variable $X$ is sub-exponential if
there exists some constant $K_1>0$ such that
$\PP(|X|>t)\leq \exp(1-t/K_1)$ for all $t\geq 0$. The sub-exponential
norm of $X$ is defined as
$\|X\|_{\psi_1}=\sup_{p\geq 1}p^{-1}(\EE|X|^p)^{1/p}$. A random
variable $X$ is sub-Gaussian if there exists some constant $K_2>0$
such that $\PP(|X|>t)\leq \exp(1-t^2/K^2_2)$ for all $t\geq 0$. The
sub-Gaussian norm of $X$ is defined as
$\|X\|_{\psi_2}=\sup_{p\geq 1}p^{-1/2}(\EE|X|^p)^{1/p}$. Denote $a\vee b=\max(a,b)$.

\section{The Proposed Methodology}
\label{seclinear}
\subsection{Setup}

Suppose that we observe a simple random sample of size $n$ from a
population of interest.  For each unit $i$, we observe a 3-tuple
$(T_i,Y_i,X_i )$ where $X_i $ is a $d$-dimensional vector of
pre-treatment covariates, $Y_i$ is an outcome variable, and $T_i$ is a
binary treatment variable denoting whether the observation receives
the treatment $(T_i=1)$ or not $(T_i=0)$.  Let $Y_i(1)$ and $Y_i(0)$
denote the potential outcomes under the treatment and control
conditions, respectively.  This notation implies the stable unit
treatment value assumption \citep{rubi:90}.  Then, the observed
outcome can be written as $Y_i = Y_i(T_i)$. Our goal is to infer the
average treatment effect (ATE),
\begin{equation}
  \mu^* \ = \ \EE\{Y_i(1) - Y_i(0)\}. \label{eq:ATE}
\end{equation}

We focus on the estimation of $\mu_1^\ast=\EE\{Y_i(1)\}$ since
$\mu_0^\ast =\EE\{Y_i(0)\}$ can be estimated in a similar manner.  We
impose a working parametric model $\pi(X_i ^\top \beta)$ for the
treatment assignment mechanism, which is known as the propensity score
$\PP(T_i=1\mid X_i)$, where $\pi(\cdot)$ is a known function and
$\beta$ is an unknown $d$-dimensional vector. In this work, we consider the
settings where the number of covariates is possibly much greater than
the sample size, i.e., $d\gg n$. When the propensity score model is
correctly specified, we have
\begin{equation}\label{eqps}
\PP(T_i = 1 \mid X_i ) \ = \ \pi(X_i ^\top \beta^*),
\end{equation}
for some $\beta^*\in\RR^d$. 
Similarly, for the outcome variable we impose a linear working
model. When the working model is correctly specified, we have
\begin{equation}\label{eqoutcome}
\EE\{Y_i(1)\mid X_i \} \ = \ K_1(X_i ),
\end{equation}
where $K_1(X_i )=\alpha^{*\top}X_i$ for some $\alpha^*\in\RR^d$. An
extension to the generalized linear models will be studied in
Section~\ref{secgeneral}. In general, the propensity score model
(\ref{eqps}) or the outcome model (\ref{eqoutcome}) can be
misspecified. We begin by assuming both models (\ref{eqps}) and
(\ref{eqoutcome}) hold. When studying theoretical properties of our
proposed methodology in Section~\ref{sec_theory}, however, we will
consider the situations, in which either model (\ref{eqps}) or
(\ref{eqoutcome}) does not hold.

\subsection{High-Dimensional Covariate Balancing Propensity
  Score}\label{secmethod}

In many applications, it is often reasonable to assume that the
propensity score model is sparse or approximately sparse. Under the 
sparsity assumption, \cite{tibshirani1996regression} and
\cite{fan2001variable} proposed the penalized maximum likelihood
estimators (PMLEs) for parameter estimation and
prediction. Unfortunately, the PMLE cannot be directly used with the
Horvitz-Thompson estimator to infer $\mu^*_1=\EE\{Y_i(1)\}$ because
the PMLE may incur a large bias due to shrinkage and its limiting
distribution is often non-normal. Thus, the resulting estimator may
have a slower rate of convergence and an intractable limiting
distribution.

To address this problem, we estimate the propensity score by
optimizing covariate balance between the treatment and control groups.
To this end, we distinguish the following two types of covariate
balancing properties.
\begin{definition}[Covariate Balancing Properties]
  Let $\hat\pi=\pi(X^\top \hat\beta)$ denote an estimator of the
   propensity score $\PP(T= 1 \mid X)$ with
  $\hat\beta$ being an estimator of $\beta^\ast$, which is the true
  value of $\beta$.
  \begin{itemize}
\item [(a)] We call $\hat\pi$ satisfies the {\it strong} covariate
  balancing property if the following equality holds,
\begin{equation}\label{eqstrong}
\sum_{i=1}^n\left(\frac{T_i}{\hat\pi_i}-1\right)X_{i} \ = \ 0.
\end{equation}
\item[(b)] We call $\hat\pi$ satisfies the {\it weak} covariate
  balancing property if the following equality holds,
\begin{equation}\label{eqweak}
\sum_{i=1}^n\left(\frac{T_i}{\hat\pi_i}-1\right)\alpha^{*\top}X_{i} \ = \ 0,
\end{equation}
where $\alpha^\ast$ is defined by $K_1(X_i )=\alpha^{*\top}X_i $ in
equation~\eqref{eqoutcome}.
\end{itemize}
\end{definition}

Although the strong covariate balancing property implies the weak one,
the converse does not necessarily hold.  The existing covariate
balancing propensity score methods aim to achieve the strong covariate
balancing property, which balances the mean of every component of
$X_i$ \citep[e.g.,][]{MR3153941,fan2016}.  However, constructing an
estimator $\hat\pi$ with the strong covariate balancing property is
difficult in high-dimensional settings.  When $d > n$, the estimator
$\hat\beta$ that satisfies equation~\eqref{eqstrong} is not unique and
therefore not even well defined. In addition, imposing additional
penalty or constraint may introduce bias because it may not satisfy
the strong covariate balancing property.

To overcome this difficulty, we propose to estimate the propensity
score such that equation~\eqref{eqweak} rather than
equation~\eqref{eqstrong} approximately holds.  We show that the weak
covariate balancing property is sufficient to remove the bias from the estimation of the propensity score model. Here, we first introduce the proposed methodology, which we
call the high-dimensional covariate balancing propensity score
(\hCBPS).
\begin{description}
\item [Step 1:] Define a generalized quasi-likelihood function as
\begin{equation}\label{eq_quasi1}
Q_n(\beta)=\frac{1}{n}\sum_{i=1}^n \int_0^{\beta^{\top}X_i }\left\{\frac{T_i}{\pi(u)}-1\right\}w_1(u) du,
\end{equation}
where $w_1(\cdot)$ is a positive weight function. Compute the regularized estimator
\begin{equation}\label{eqhatbeta}
\hat\beta \ = \ \argmin_{\beta\in\RR^d} -Q_n(\beta)+\lambda\|\beta\|_1,
\end{equation}
where $\lambda>0$ is a tuning parameter. 
\item [Step 2:] Define a weighted least square loss function using the treatment group as 
\begin{equation}\label{eq_loss1}
L_n(\alpha)=\frac{1}{n}\sum_{i=1}^n T_i w_2(\hat\beta^\top X_i)(Y_i-\alpha^{\top}X_i )^2,
\end{equation}
where $w_2(\cdot)$ is another positive weight function. Compute the regularized estimator
\begin{equation}\label{eqhatalpha}
\tilde\alpha \ = \ \argmin_{\alpha\in\RR^d}L_n(\alpha)+\lambda'\|\alpha\|_1,
\end{equation}
where $\lambda'>0$ is a tuning parameter.
\item [Step 3:] Let $\tilde S=\{j: |\tilde\alpha_j|>0\}$ denote the
  support of $\tilde\alpha$ and $X_{\tilde S}$ denote the
  corresponding subset of $X$. We calibrate the initial estimator
  $\hat\beta_{\tilde{S}}$ to balance $X_{\tilde{S}}$.
  Specifically, we solve,
\begin{equation}\label{eqcovariatebalance}
  \tilde\gamma \ = \ \argmin_{\gamma\in\RR^{|\tilde
      S|}}\|g_n(\gamma)\|^2_2 ~~\textrm{where}~~ g_n(\gamma) \ = \ \frac{1}{n}\sum_{i=1}^n \left\{\frac{T_i}{\pi(\gamma^\top X_{i\tilde S}+\hat\beta^\top_{\tilde S^c}X_{i\tilde S^c})}-1\right\}X_{i \tilde S}
\end{equation}
We then set $\tilde\beta=(\tilde\gamma, \hat\beta_{\tilde S^c})$
and $\tilde\pi_i=\pi(\tilde\beta^\top X_i)$.

\item [Step 4:] Estimate $\mu_1^\ast=\EE\{Y_i(1)\}$ by the Horvitz-Thompson estimator
  $\hat \mu_1=\frac{1}{n} \sum_{i=1}^{n}T_iY_i/\tilde\pi_i$.
\end{description}

In Step~1, we obtain an initial estimate of the propensity score via
the penalized M-estimation approach. We refer to the function
$Q_n(\beta)$ as the generalized quasi-likelihood function, as its
construction is similar to the quasi-likelihood function for
generalized linear models \citep{wedderburn1974quasi}. To understand
how the generalized quasi-likelihood function is motivated, we compute the corresponding quasi-score function,
\begin{equation}\label{eq_quasi2}
\frac{\partial Q_n(\beta)}{\partial \beta} \ = \ \frac{1}{n}\sum_{i=1}^n \left\{\frac{T_i}{\pi(\beta^{\top}X_i)}-1\right\}w_1(\beta^{\top}X_i)X_i. 
\end{equation}
Since \eqref{eq_quasi2} is an unbiased estimating function for
$\beta$, $Q_n(\beta)$ serves as a legitimate quasi-likelihood function
that integrates the quasi-score function \eqref{eq_quasi2}. The
quasi-likelihood function $Q_n(\beta)$ depends on the choice of
weighting function $w_1(u)$. In particular, we consider the following
two examples.
\begin{itemize}
\item[(a)] If $w_1(u)=\pi(u)$, \eqref{eq_quasi2} is identical to the
  score function for the logistic regression and thus $Q_n(\beta)$
  reduces to the standard quasi-likelihood function for the treatment
  variable.
\item[(b)] If $w_1(u)=1$, the quasi-score function \eqref{eq_quasi2}
  leads to the strong covariate balancing equation
  (\ref{eqstrong}). Consequently, we call $Q_n(\beta)$ with $w_1(u)=1$
  as the covariate balancing loss function.
\end{itemize}
Thus, in Step 1, we allow a broad class of initial estimators
$\hat\beta$, including the penalized (quasi)-maximum likelihood
estimator and many other penalized M-estimators corresponding to
different $w_1(u)$.
By computing the Hessian matrix of $Q_n(\beta)$, we find that
(\ref{eqhatbeta}) can be a non-convex optimization problem depending
on the choice of $w_1(u)$. The non-convexity may pose computational
challenges. For instance, the gradient descent algorithm can be
trapped at a local solution which is far from the global maximizer. To
avoid the computational issue, we mainly focus on the concave
quasi-likelihood function $Q_n(\beta)$. It is easy to verify that
$Q_n(\beta)$ with $w_1(u)=\pi(u)$ in case (a) and $w_1(u)=1$ in case
(b) are both concave.

In Step~2, we fit the outcome model using a class of penalized
weighted least square estimators. We allow the weight
$w_2(\hat\beta^\top X_i)$ to depend on $X_i$ and also the initial
estimator $\hat\beta$ from Step 1. For instance, we have the following
examples.
\begin{itemize}
\item[(a')] If $w_2(u)=1$, $L_n(\alpha)$ is the classical least square
  loss function in the treatment group.
\item[(b')] If $w_2(u)=1/\pi(u)$, $L_n(\alpha)$ is known as the
  inverse propensity score weighted least square loss.
\item[(c')] If $w_2(u)=\pi'(u)/\pi^2(u)$, $L_n(\alpha)$ remains a
  valid loss function for estimating $\alpha$. It is shown in Section
  \ref{secrobust2} that this loss function plays an important role
  when studying the robustness of the proposed estimator to
  misspecified outcome models. In the following, we call this loss
  function as the propensity score adjusted least square loss.
\end{itemize}


Step~3 removes the bias induced by the penalized estimators used in
Steps~1~and~2.  We calibrate the estimated propensity score by
balancing a subset of covariates $X_{\tilde S}$, which represent the
variables selected for the outcome model.
Equation~\eqref{eqcovariatebalance} implies that the proposed \hCBPS\
methodology achieves the strong covariate balancing property only for
these covariates $X_{\tilde S}$ but not for the other covariates
$X_{\tilde S^c}$.  Thus, unlike the original \CBPS\ methodology, the
\hCBPS\ methodology does not achieve the strong covariate balancing
property.  Interestingly, however, the \hCBPS\ methodology does
approximately satisfy the weak covariate balancing property if
$\alpha^*$ can be well approximated by $\tilde\alpha$. Specifically,
we have
\begin{equation}
\sum_{i=1}^n\left(\frac{T_i}{\tilde\pi_i}-1\right)\alpha^{*\top}X_{i}\ \approx \ \sum_{i=1}^n\left(\frac{T_i}{\tilde\pi_i}-1\right)\tilde\alpha^{\top}X_{i}\ = \ \sum_{i=1}^n\left(\frac{T_i}{\tilde\pi_i}-1\right)\tilde\alpha^{\top}_{\tilde S}X_{i \tilde S}\ = \ 0,\label{eqbalance}
\end{equation}
where the first equality follows from $\tilde\alpha_{\tilde S^c}=0$
and the second equality holds due to
equation~\eqref{eqcovariatebalance}.

In Step 4, we estimate $\mu^*_1$ using the Horvitz-Thompson
estimator. In the following, we comment on the connection between the
proposed estimator and the other commonly used estimators. First, our
estimator can be written as the Horvitz-Thompson estimator with the
normalized weights, which is known as the Hajek estimator
\citep{haje:71},
$$
\hat \mu_1 \ = \ \frac{1}{n} \sum_{i=1}^{n} \frac{T_iY_i}{\tilde\pi_i}
\ = \ \frac{\sum_{i=1}^{n} T_iY_i/\tilde\pi_i}{\sum_{i=1}^{n} T_i/\tilde\pi_i},
$$
The second equality follows because
$\sum_{i=1}^{n}(T_i/\tilde\pi_i-1)/n=0$ so long as an intercept is
included in $X_{i\tilde S}$. \cite{imbens2005mean} and
\cite{busso2014new} showed that the normalized Horvitz-Thompson
estimator tends to be more stable than the unnormalized version
numerically. Thus, we expect that the proposed estimator has a better
finite sample performance than the standard (i.e., unnormalized)
Horvitz-Thompson estimator.

Second, our estimator can be also rewritten as an AIPW estimator with
the linear outcome model \citep{robins1994estimation},
\begin{equation}\label{eqAIPW}
\hat \mu_1 \ = \ \frac{1}{n}\sum_{i=1}^n \frac{T_iY_i}{\tilde\pi_i} \
= \ \frac{1}{n}\sum_{i=1}^n
\frac{T_iY_i}{\tilde\pi_i}+\frac{1}{n}\sum_{i=1}^n
\left(1-\frac{T_i}{\tilde\pi_i}\right) \tilde\alpha^\top X_i
\end{equation}
where the second equality follows from two equalities
in~\eqref{eqbalance}.  We conduct a further technical comparison with
the AIPW estimator in Section~\ref{seccompare}.

From a practical perspective, \cite{rubi:08} advocated an outcome free
design for the treatment effect estimation, in which the propensity
score is estimated without reference to the outcome information in a
similar spirit to the randomized experiment.  We note that our
estimator does not fall into the framework of the outcome free design,
because we recalibrate the propensity score using the outcome
information in Step 3.

Our procedure differs from the existing methods on high-dimensional
regressions; see \cite{zhang2011confidence},
\cite{javanmard2013confidence}, \cite{van2013asymptotically},
\cite{belloni2013honest}, \cite{ning2014general},
\cite{cai2015confidence}, and \cite{dukes2018high}, among many
others. The main idea of these methods is to correct the bias of the
Lasso-type estimators or the score function by inverting the optimality condition or projecting to the tangent space of the nuisance function. 
In contrast, we remove the bias of the Lasso estimators $\hat\beta$ and
$\tilde\alpha$ using a covariate balancing strategy.



\section{Theoretical Properties of the Proposed Estimator}\label{sec_theory}
We now study the theoretical properties of the proposed estimator.  We begin by presenting the required assumptions.

\subsection{Assumptions}\label{sec_ass}

\begin{assumption}[Unconfoundedness]\label{assun}
  The treatment assignment is unconfounded, i.e.,
  $\{Y_i(1), Y_i(0)\} \ \perp \ T_i \mid X_i $.
\end{assumption}

\begin{assumption}[Overlap]\label{asspropen}
  There exists a constant $c_0>0$ such that $\pi_i^*\geq c_0$ for
  $1\leq i\leq n$, where $\pi_i^*=\pi(X_i^\top\beta^*)$
\end{assumption}

Assumption \ref{assun} implies that there is no unmeasured confounders
while Assumption \ref{asspropen} requires that all samples have a
positive probability to receive the treatment.  Together, these
represent the standard strong ignorability condition common to
propensity score methods \citep{rose:rubi:83}; see e.g., Assumption~1
of \cite{farrell2015robust} and Assumption~6 of
\cite{athey2016efficient}. To estimate the treatment effect, one also
needs to identify $\EE\{Y(0)\}$, which requires a similar overlap
assumption $\pi_i^*\leq 1-c_1$ for some constant $c_1>0$.

\begin{assumption}[Sub-Gaussian condition] \label{asssubg} Assume that
  $\epsilon_1=Y(1)-\alpha^{*\top}X$ and $X_j$ satisfy
  $\|\epsilon_1\|_{\psi_2}\leq C_\epsilon$ and
  $\|X_j\|_{\psi_2}\leq C_X$ for any $1\leq j\leq d$, where $C_X$ and
  $C_{\epsilon}$ are positive constants. 
\end{assumption}	

Assumption \ref{asssubg} controls the tail behavior of the error
$\epsilon_1$ and the covariate $X_j$, which facilitates the use of
many existing concentration inequalities in high-dimensional
statistics. Similar sub-Gaussian conditions are imposed by
\cite{athey2016efficient} in their Theorem 5.
\cite{belloni2013program, farrell2015robust} relaxed the sub-Gaussian
condition on $\epsilon_1$ to the bounded $q$th moment for some $q>4$
under a slightly stronger sparsity assumption than our sparsity
assumption below. 

\begin{assumption}[Sparsity]\label{asssparse}
Assume that $(s_1\vee s_2)\log (d\vee n)/n^{1/2}=o(1)$ as $s_1, s_2, d, n\rightarrow\infty$, where $s_1=\|\beta^*\|_0$ and $s_2=\|\alpha^*\|_0$. Recall that $a\vee b=\max(a,b)$. 
\end{assumption}

Assumption \ref{asssparse} requires that the propensity score model
and outcome model are sparse. Since we consider the high-dimensional
case with $d\gg n$, the sparsity assumption plays an important role in
the regularized M-estimation of the propensity score model and outcome
model. In particular, if $s_1\asymp s_2\asymp n^\kappa$ for some
$\kappa<1/2$, then the condition reduces to
$d=o(\exp(n^{1/2-\kappa}))$. This condition is similar to that in
\citet{belloni2013program,belloni2014inference} and
\citet{farrell2015robust}, where they imposed a slightly stronger
condition with $\log (d\vee n)$ replaced by $\{\log (d\vee n)\}^q$ for
some $q>1$.

This sparsity assumption will be further relaxed later in the
paper. To preview these results, we note that Remark \ref{remlinear}
of Section \ref{sec_correct} discusses the approximate sparsity
assumption and Remark \ref{rem1} of Section \ref{sec_correct}
considers a simple modification of the algorithm based on the sample
splitting approach \citep{Chernozhukov2016}, which requires a weaker
sparsity assumption $(s_1 s_2)^{1/2}\log (d\vee n)/n^{1/2}=o(1)$.

\begin{assumption}[Eigenvalue condition]\label{asseigen}
  Denote $\Sigma=\EE(X^{\otimes 2})$. There exists a constant $C>0$
  such that
  $C\leq
  \lambda_{\min}(\Sigma_{SS})\leq\lambda_{\max}(\Sigma_{SS})\leq
  1/C$ for any $S\subset\{1,...,d\}$ with $|S|\lesssim (s_1 \vee s_2)\log n$. 
\end{assumption}

When the dimension $d$ is fixed, this assumption simply requires that
the design matrix has full column rank, which is a standard regularity
condition for regression problems.  In high dimension, Assumption
\ref{asseigen} implies the well known sparse eigenvalue condition
introduced by \cite{bickel2009simultaneous} to study the Lasso
estimator; see Lemma 1 in \cite{belloni2013least}. The same sparse
eigenvalue condition is imposed by \cite{belloni2014inference}. We
also refer to Assumption~5 of \cite{athey2016efficient} and Section
6.2 of \cite{farrell2015robust} for a similar restricted eigenvalue condition. Since our assumption only applies to any $S$ by $S$
submatrix of $\Sigma$, it is weaker than
$C\leq \lambda_{\min}(\Sigma)\leq\lambda_{\max}(\Sigma)\leq 1/C$,
imposed by \cite{van2013asymptotically}, \cite{ning2014general}, and
\cite{cai2015confidence} for high-dimensional inference.

\begin{assumption}[Propensity score and weight functions]\label{assweight}
Assume that $Q_n(\beta)$ is a concave function. Let $C,C'$ denote positive constants, which may change from line to line. 
\begin{itemize}
\item[(1)] The propensity score model $\pi(u)$ satisfies  $C\leq (\pi_i^{*})'\leq 1/C$, and there exist constants $r>0$ and $C'>0$ such that the Lipschitz condition holds locally, i.e., $|\pi'(u)-\pi'(v)|\leq C'|u-v|$ for any $u,v\in [X_i^\top\beta^*-r, X_i^\top\beta^*+r]$ and $1\leq i\leq n$. 
\item[(2)] The weight $w_1(u)$ satisfies $C\leq w_{1i}^{*}\leq 1/C$, $(w_{1i}^{*})'\leq 1/C$, and the local Lipschitz condition $|w_1'(u)-w_1'(v)|\leq C'|u-v|$ for any $u,v\in [X_i^\top\beta^*-r, X_i^\top\beta^*+r]$ and $1\leq i\leq n$, where $w_{1i}^*=w_1(X_i^\top\beta^*)$. 
\item[(3)] The weight $w_2(u)$ satisfies $C\leq w_{2i}^{*}\leq 1/C$ and $(w_{2i}^{*})'\leq 1/C$ for $1\leq i\leq n$,  where $w_{2i}^*=w_2(X_i^\top\beta^*)$. Assume $w'_2(u)$ is continuous. 
\end{itemize}
\end{assumption}

Finally, Assumption \ref{assweight} imposes mild regularity conditions
on the propensity score function and weight functions. In part (1), we
assume $\pi(u)$ is differentiable and its derivative is bounded and
Lipschitz around $X_i^\top\beta^*$. Under the overlap assumption
$c_1\leq \pi_i^*\leq 1-c_1$, part (1) holds for the logistic
regression without any further conditions. In part (2) and (3), we
assume mild conditions on the magnitude and smoothness of $w_1(u)$ and
$w_2(u)$. Again, if $\pi(u)$ is the logistic function and the overlap
assumption holds, all examples of $w_1(u)$ and $w_2(u)$ discussed in
Section \ref{secmethod} satisfy the regularity conditions in part (2)
and (3). Thus, Assumption \ref{assweight} holds for the
logistic propensity score model without any further conditions.


\subsection{Asymptotic Distribution under Correct Model Specification}\label{sec_correct}

In this subsection, we derive the theoretical results for the proposed
\hCBPS\ estimator $\hat\mu_1$ when both the propensity score
model~\eqref{eqps} and the outcome model~\eqref{eqoutcome} are
correctly specified. Recall that our estimator $\hat\mu_1$ depends on
the choice of the two weight functions, i.e., $w_1(u)$ in Step 1 and
$w_2(u)$ in Step 2. In the following, we establish the asymptotic
normality and semiparametric efficiency of $\hat\mu_1$ for any weight
function $w_1(u)$ and $w_2(u)$.


\begin{theorem}[Asymptotic Normality and Semiparametric Efficiency]\label{thm1}
  Suppose that both the propensity score model~\eqref{eqps} and the
  outcome model~\eqref{eqoutcome} are correctly specified and
  Assumptions 1--6 hold.  If we take
  $\lambda\asymp\lambda'\asymp \{{\log (d\vee n)/n}\}^{1/2}$, then the
  estimator $\hat{\mu}_1$ with any weight function $w_1(u)$ and
  $w_2(u)$ satisfies
$$
\hat{\mu}_1-\mu_1^* \ = \ \frac{1}{n}\sum_{i=1}^n\left[\frac{T_i}{\pi^*_i}\{Y_i(1)-\alpha^{*\top}X_i \}+\alpha^{*\top}X_i -\mu_1^*\right]+O_p\left(\frac{(s_1\vee s_2)\log (d\vee n)}{n}\right),
$$
as $s_1, s_2, d, n\rightarrow\infty$. Let $V$ be the semiparametric asymptotic variance bound, i.e., 
$$
V \ = \ \EE\left\{\frac{1}{\pi^*}\EE(\epsilon_1^2\mid X)+(\alpha^{*\top}X-\mu_1^*)^2\right\}.
$$ 
Assume that $\EE(\epsilon_1^2\mid X)\geq c$ for some constant $c>0$ and $\EE(\alpha^{*\top} X)^4=O(s_2^2)$. Then, $n^{1/2}(\hat{\mu}_1-\mu_1^*)/V^{1/2}\rightarrow_d N(0,1)$.
\end{theorem}

The theorem shows that $\hat \mu_1-\mu_1^*$ is asymptotically
equivalent to the average of the efficient score functions and hence
$\hat \mu_1$ is locally efficient under the correct model
specification. In addition, the asymptotic distribution of
$\hat \mu_1$ does not depend on choice of the weight functions
$w_1(u)$ and $w_2(u)$, provided that they satisfy
Assumption~\ref{assweight}. The intuition is that, as long as the weak
covariate balancing property is approximately attained, the choice of
the weight functions in the first two steps is less important.

To prove the asymptotic normality of $\hat\mu_1$, we further require
that the variance of the noise cannot tend to 0, i.e.,
$\EE(\epsilon_1^2\mid X)\geq c>0$. This guarantees the non-degeneracy
of the asymptotic variance $V$. We also assume
$\EE(\alpha^{*\top} X)^4=O(s_2^2)$ in order to verify the Lyaponov
condition for the central limit theorem. This is a mild technical
condition. For instance, if $X$ is a sub-Gaussian vector and
$\|\alpha^*\|_2=O(s_2^{1/2})$, then
$\|\alpha^{*\top} X\|_{\psi_2}\leq
\|\alpha^*\|_2\|X\|_{\psi_2}=O(s_2^{1/2})$.  This further implies the
desired condition $\EE(\alpha^{*\top} X)^4=O(s_2^2)$ by the definition
of the sub-Gaussian norm.

We note that the asymptotic variance $V$ depends on the true data
generating process, which is allowed to change with $d$ and also
$n$. For this reason, we consider the limiting distribution of the
standardized statistic $n^{1/2}(\hat{\mu}_1-\mu_1^*)/V^{1/2}$ as
$n,d\rightarrow\infty$. \cite{hahn1998role} proved that $V$ is the
semiparametric asymptotic variance bound, when both the propensity
score and outcome models are treated as nuisance.  He further proposed
a nonparametric IPW estimator for fixed $d$ that attains this
semiparametric efficiency bound.  We show that, when the
high-dimensional models~\eqref{eqps}~and~\eqref{eqoutcome} are both
correctly specified, the estimator $\hat \mu_1$ attains the same bound
and hence locally efficient.

Our variance bound $V$ is different from the ``oracle efficiency
bound", which is the semiparametric variance bound with the known
support of the propensity score and outcome models
\citep{hahn2004functional}. Since the support of both models is
unknown and the variable selection consistency does not hold under our
assumptions, the estimation of the support set leads to additional
uncertainty. This explains why our method cannot attain the oracle
efficiency bound. We refer to Section 5.3 of \cite{farrell2015robust}
for further discussion on this point.

Since our goal is to estimate the causal effects rather than the
coefficients in the propensity score and outcome models, we show that
the asymptotic normality of $\hat \mu_1$ does not rely on the variable
selection consistency in either model. It is known that variable
selection consistency requires more stringent conditions, e.g., signal
strength condition and irrepresentable condition
\citep{zhao2006model}. Theorem~\ref{thm1} does not require these
conditions.

\begin{remark}[Approximate Sparsity]\label{remlinear}
  Theorem~\ref{thm1} assumes that the propensity score and outcome
  models are sparse.  However, the same conclusion holds for the class
  of approximately sparse models. Specifically, assume that
$$
\EE(Y_i(1)\mid X_i) =X_i^{\top} \alpha^{*}  + r_i, ~~\textrm{and}~~\PP(T_i = 1 \mid X_i ) \ = \ \pi(X_i ^\top \beta^*+u_i),
$$ 
where $s_1=\|\beta^*\|_0$ and $s_2=\|\alpha^*\|_0$ and $r_i, u_i$ are
the approximation errors. By introducing $r_i, u_i$ in these models,
we allow for the nonlinear effect of $X_i$ and the non-sparse effect
due to weak signals in the models. Using a proof similar to the one
for the theorem, we can show that Theorem~\ref{thm1} holds so long as
the approximation errors satisfy
$$
\sum_{i=1}^n r^2_i=O(s_2),~\sum_{i=1}^n u^2_i=O(s_1)~~\textrm{and}~~\sum_{i=1}^n r_iu_i=o(n^{1/2}).
$$
Thus, our results are robust to the minor violations of the linearity
and sparsity assumptions.
\end{remark}

\begin{remark}[Sample Splitting]\label{rem1}
  In a recent work, \cite{Chernozhukov2016} proposed a double machine
  learning method based on the sample splitting technique to relax the
  sparsity assumption. In the supplementary materials, we proposed a
  modified algorithm based on the sample splitting, so that Assumption~\ref{asssparse} is relaxed
  to a weaker assumption $(s_1 s_2)^{1/2}\log (d\vee
  n)/n^{1/2}=o(1)$. Ignoring the logarithmic factors of $d$ and $n$,
  Assumption~\ref{asssparse} requires $s_1=o(n^{1/2})$ and
  $s_2=o(n^{1/2})$. In contrast, by using the sample splitting
  technique, we only require a weaker condition $s_1s_2=o(n)$, which
  may still hold if one model is dense (e.g., $n^{1/2}\ll s_1\ll n$)
  and the other model is sufficiently sparse (e.g., $s_2\ll
  n^{1/2}$). However, the sample splitting method incurs further
  computational cost and may not be stable when the sample size is
  relatively small.
\end{remark}

\begin{remark}[Sample Boundedness]\label{remsample}
  Unlike many of the existing estimators, the proposed \hCBPS\
 method guarantees that $\hat \mu_1$ lies in the range
  of $\{Y_i: T_i=1, i=1,...,n\}$. This sample boundedness property
  \citep{robins2007comment} holds because, by construction, the
  covariate balancing equation satisfies
\begin{equation}\label{eqsample}
\frac{1}{n}\sum_{i=1}^n \left(\frac{T_i}{\tilde\pi_i}-1\right) \ = \ 0,
\end{equation}
so long as an intercept is included in $X_{i\tilde
  S}$. Equation~\eqref{eqsample} implies that the estimated propensity
score $\tilde\pi_i$ must be greater than or equal to $1/n$ for any
treated observation. In contrast, the estimated propensity score
$\pi_i^*$ for the $i$th observation via the penalized maximum
likelihood estimation can become very close to $0$, leading to
extremely large weights for some observations and unstable causal
effect estimates.  To see why the sample boundedness property holds,
we have
\begin{equation*}
\frac{1}{n}\sum_{i=1}^n \frac{T_iY_i}{\tilde\pi_i}\ \geq \ \frac{\min_{i: T_i=1} Y_i}{n}\sum_{i=1}^n \frac{T_i}{\tilde\pi_i}\ =  \ \min_{i: T_i=1} Y_i,
\end{equation*}
where the last equality follows from
equation~\eqref{eqsample}. Similarly, we can show that
$\hat\mu_1\leq \max_{i: T_i=1} Y_i$. 
\end{remark}

Finally, to construct a confidence interval for $\mu_1^*$, we estimate
$V$ by
\begin{equation}\label{eqestvar}
  \hat V \ = \ \frac{1}{n}\sum_{i=1}^n \left\{\frac{T_i}{\tilde\pi_i^2}(Y_i-\tilde\alpha^\top X_i)^2+(\tilde\alpha^\top X_i -\hat\mu_1)^2\right\}.
  \end{equation}
The following corollary shows that $\hat V$ is a consistent estimator of $V$ and therefore we obtain valid confidence intervals for $\mu_1^*$. 

\begin{corollary}[Honest Confidence Intervals] \label{corCI} Suppose
  that the assumptions in Theorem~\ref{thm1} hold. Then,
$$
|\hat V-V| \ = \ O_p\left((s_1\vee s_2)\sqrt{\frac{{\log (d\vee n)}}{{n}}}\right).
$$
Given $0<\eta\leq 1$, define the $(1-\eta)$-confidence interval as $\mathcal{I} \ = \ (\hat\mu_1-z_{1-\eta/2}({\hat V/n})^{1/2},\
  \hat\mu_1+z_{1-\eta/2}({\hat V/n})^{1/2})$, where $z_{1-\eta/2}$ is the $(1-\eta/2)$ quantile of a standard normal distribution.  Then, 
\begin{equation}\label{thm1unif2}
\Big|\PP(\mu_1^*\in\mathcal{I})-(1-\eta)\Big| =o(1).
\end{equation}

\end{corollary}
Indeed, this confidence interval $\mathcal{I}$ is honest in the sense that
equation~\eqref{thm1unif2} holds uniformly over all probability
distributions that satisfy Assumptions 1-6. In addition, the proof of
Corollary \ref{corCI} holds even if the error $\epsilon_1$ is
heteroskedastic, i.e., $\EE(\epsilon_1^2\mid X)$ depends on the value
of $X$.

In the Supplementary Materials, we further extend these theoretical
results to the estimation of the average treatment effect for the
treated (ATT). 

\subsection{Asymptotic Distribution under Misspecified Propensity Score
  Models}
\label{secrobust}

We next investigate the robustness of the proposed \hCBPS\ methodology
to the misspecification of propensity score model. In this subsection,
we assume that the true propensity score $\pi^*=\PP(T=1\mid X)$ does
not belong to the assumed parametric class
$\{\pi( X ^\top \beta): \beta\in\RR^d\}$. To study the limiting
behavior of our estimator $\hat\mu_1$ in this setting, we first define
the estimand of $\hat\beta$ in Step 1. Given the generalized
quasi-likelihood function $Q_n(\beta)$, the estimand of $\hat\beta$
in~\eqref{eqhatbeta} is defined as
$$
\beta^o \ = \ \argmax \EE\left[\int_0^{\beta^{\top}X_i }\left\{\frac{T_i}{\pi(u)}-1\right\}w_1(u) du\right].
$$
We note that the estimand $\beta^o$ implicitly depends on the choice
of the weight function $w_1(u)$, and when the model is correctly
specified, $\beta^o$ reduces to $\beta^*$. In the following
proposition, we assume that the estimand $\beta^o$ is sparse, which is
a technical assumption required to study misspecified models in
high-dimensional settings \citep{buhlmann2015high}. For instance, under this
assumption, it can be shown that $\beta^o$ can be consistently
estimated by $\hat\beta$.  Without similar assumptions, the high-dimensional parameter $\beta^o$ may not
be estimable. 
The following proposition establishes the
asymptotic properties of $\hat\mu_1$ under misspecified propensity
score models.

\begin{proposition} {\sc (Consistency and Asymptotic Normality under
    Misspecified Propensity Score Models)} \label{thmmis} Suppose that
  the outcome model~\eqref{eqoutcome} is correctly specified, but the
  propensity score model~\eqref{eqps} is misspecified. Assumptions 1-6
  hold with $\beta^*$ replaced by $\beta^o$.  If we take
  $\lambda\asymp\lambda'\asymp \{{\log (d\vee n)/n}\}^{1/2}$, then the
  estimator $\hat{\mu}_1$ with any weight functions $w_1(u)$ and
  $w_2(u)$ satisfies
\begin{equation}\label{eqthmmis1}
\hat\mu_1 - \mu_1^*\ = \ O_p\left(\sqrt{\frac{(s_1\vee s_2) \log (d\vee n)}{n}}\right).
\end{equation}  
Moreover, if we set $w_1(u)=1$, then for any $w_2(u)$ we have
$$
\hat{\mu}_1-\mu_1^* \ = \
\frac{1}{n}\sum_{i=1}^n\left\{\frac{T_i}{\pi^o_i}(Y_i(1)-\alpha^{*\top}
  X _i)+\alpha^{*\top} X _i-\mu_1^*\right\}+O_p\left(\frac{(s_1\vee
    s_2)\log (d\vee n)}{n}\right), 
$$
where $\pi_i^o=\pi(X_i^\top\beta^o)$. Assume that
$\EE(\epsilon_1^2\mid X)\geq c$ for some constant $c>0$ and
$\EE(\alpha^{*\top} X)^4=O(s_2^2)$. This implies
$n^{1/2}(\hat{\mu}_1-\mu_1^*)/V_{\sf mis-ps}^{1/2}\rightarrow_d N(0,1)$
where
$$
V_{\sf mis-ps} \ = \ \EE\left\{\frac{\pi^*}{\pi^{o2}}\EE(\epsilon_1^2\mid  X )+(\alpha^{*\top} X -\mu_1^*)^2\right\}.
$$ 
\end{proposition}

This proposition shows that our estimator $\hat\mu_1$ remains
consistent, but the convergence rate in (\ref{eqthmmis1}) can be slower
than $n^{-1/2}$. When the dimension $d$ is fixed, (\ref{eqthmmis1})
reduces to $\hat\mu_1 - \mu_1^*=O_p((\log n/n)^{1/2})$, which agrees
with the convergence rate of the doubly robust estimators in low
dimension up to a $\log n$ factor, see
\cite{robins1994estimation,bang2005doubly,robins2007comment,cao2009improving,tan2010bounded,van2011targeted,vermeulen2015bias}. However,
in high dimension, the convergence rate becomes slower because of the
extra bias in the propensity score estimation when the model is
misspecified.

The second part of this proposition states that, if we carefully choose
the weight function $w_1(u)=1$ in the algorithm, the
root-$n$ consistency and asymptotic normality of $\hat\mu_1$ are
restored without paying any price. Recall that with $w_1(u)=1$, the
quasi-score function (\ref{eq_quasi2}) reduces to the strong covariate
balancing equation (\ref{eqstrong}) as in the example (b). This is the
key ingredient to remove the extra bias term in the asymptotic
expansion of $\hat{\mu}_1$ when the propensity score model is
misspecified. 

To construct the confidence interval for $\mu_1^*$, we need to
estimate $V_{\sf mis-ps}$. By inspecting the proof of
Corollary~\ref{corCI}, we can show that the estimator $\hat V$ defined
in~\eqref{eqestvar} is still consistent for $V_{\sf mis-ps}$, even if
the propensity score model is misspecified. Thus, the confidence
interval shown in Corollary~\ref{corCI} is valid whether or not the
propensity score model is misspecified. Finally, we note that if the
propensity score model is correctly specified, i.e.,
$\pi^o_i=\pi^*_i$, then the asymptotic variance $V_{\sf mis-ps}$ in
Proposition~\ref{thmmis} reduces to the asymptotic variance $V$ in
Theorem~\ref{thm1}.

\subsection{Asymptotic Distribution under Misspecified Outcome
  Models}
\label{secrobust2}

In this subsection, we study the robustness of the proposed \hCBPS\
methodology to the misspecification of the outcome model. Assume that
the propensity score model is correct, but the true conditional mean
function $\EE(Y_i(1)\mid X_i=x)$ is nonlinear in $x$, i.e., there does
not exist $\alpha^*$ such that
$\EE(Y_i(1) \mid X_i=x)=\alpha^{*\top}x$. Similar to
Section~\ref{secrobust}, we define the estimand of $\tilde\alpha$ in
Step 2 as
$$
\alpha^o \ = \ \argmin \EE\left\{T_i w_2(\beta^{*\top}X_i)(Y_i-\alpha^{\top}X_i )^2\right\},
$$
which in turn depends on the weight function
$w_2(\beta^{*\top}X_i)$. The following proposition establishes the
asymptotic properties of $\hat\mu_1$ under misspecified outcome
models.

\begin{proposition} {\sc (Consistency and Asymptotic Normality under
    Misspecified Outcome Models)}\label{thmmis2} 
  Suppose that the propensity score model~\eqref{eqps} is correctly
  specified, but the outcome model~\eqref{eqoutcome} is
  misspecified. Assumptions 1-6 hold with $\alpha^*$ replaced by
  $\alpha^o$.  If we take
  $\lambda\asymp\lambda'\asymp \{{\log (d\vee n)/n}\}^{1/2}$, then the
  estimator $\hat{\mu}_1$ with any weight functions $w_1(u)$ and
  $w_2(u)$ satisfies
\begin{equation}\label{eqthmmis2}
\hat\mu_1 - \mu_1^* \ = \ O_p\left(\sqrt{\frac{(s_1\vee s_2)\log (d\vee n)}{n}}\right).
\end{equation}  
Moreover, if we set $w_2(u)=\pi'(u)/\pi^2(u)$, then for any $w_1(u)$ we have
$$
\hat{\mu}_1-\mu_1^* \ = \
\frac{1}{n}\sum_{i=1}^n\left\{\frac{T_i}{\pi^*_i}(Y_i(1)-\alpha^{o\top}
  X _i)+\alpha^{o\top} X _i-\mu_1^*\right\}+O_p\left(\frac{(s_1\vee
    s_2)\log (d\vee n)}{n}\right). 
$$
Assume that $\EE(\epsilon_1^{o2}\mid X)\geq c$ for some constant $c>0$
and $\EE(\alpha^{o\top} X)^4=O(s_2^2)$, where
$\epsilon_1^o=Y(1)-\alpha^{o\top}X$. This implies
$n^{1/2}(\hat{\mu}_1-\mu_1^*)/V_{\sf mis-o}^{1/2}
\rightarrow_d N(0,1)$,  where
$$
V_{\sf mis-o} \ = \ \EE\left\{\frac{1}{\pi^{*}}\EE(\epsilon^{o2}\mid  X )+(\alpha^{o\top} X -\mu_1^*)^2\right\}.
$$ 
\end{proposition}

The results in this proposition are parallel to those in
Proposition~\ref{thmmis}. Specifically, when the outcome model is
misspecified, our estimator is still consistent, but has a slower
convergence rate as shown in (\ref{eqthmmis2}). However, as long as we
choose $w_2(u)=\pi'(u)/\pi^2(u)$ or equivalently the propensity score
adjusted least square loss $L_n(\alpha)$ in example~(c') of
Section~\ref{secmethod}, the desired properties such as the root-$n$
consistency and asymptotic normality are restored.  The form of the
$w_2(u)$ is designed to eliminate the bias term in the expansion of
$\hat\mu_1$ under the misspecified outcome model. Finally, the
estimator $\hat V$ in~\eqref{eqestvar} is consistent for the
asymptotic variance $V_{\sf mis-o}$.

\begin{remark}[Double Robustness and Honest Confidence
  Intervals]\label{remrobust}
  Propositions~\ref{thmmis}~and~\ref{thmmis2} together imply that our
  estimator $\hat\mu_1$ is root-$n$ consistent and asymptotically
  normal provided either the propensity score model or outcome model
  is correctly specified. This estimator does not require to know
  which of the two models is correct.  Since $\hat V$ is always
  consistent, the same confidence interval defined in Corollary
  \ref{corCI} is valid as long as one of the two models is correctly
  specified. Thus, we recommend the use of this estimator and the
  associated confidence interval in practice. Finally, we comment that
  when $w_1(u)=1$ and $w_2(u)=\pi'(u)/\pi^2(u)$, the gradients of
  $Q_n(\beta)$ and $L_n(\alpha)$ reduce to the estimating
  equations proposed by \cite{robins2007comment}.
\end{remark}

\subsection{Comparison with the Related Work}\label{seccompare}

In this subsection, we compare our method with the related
work. First, we comment on the theoretical results of the AIPW
estimator \citep{belloni2013program,farrell2015robust} and double
selection estimator \citep{belloni2014inference} when both the
propensity score and outcome models are correctly specified. Second,
we compare the results when one of the two models is
misspecified. Finally, we consider the more recent work by
\cite{athey2016efficient}, \cite{zhao2016covariate} and \cite{tan2017regularized,tan2018model}.

When both the propensity score and outcome models are correctly
specified, \citet{belloni2013program} and \citet{farrell2015robust}
showed that the AIPW estimator is asymptotically normal and efficient
in high dimension. Their assumptions and main results are parallel to
our Theorem~\ref{thm1}. However, the sample boundedness property in
Remark \ref{remsample} does not hold for the AIPW estimator in
general. We note that the above work and our Theorem \ref{thm1} can be
viewed as an extension of the semiparametric efficiency property of
the doubly robust estimators; see
\cite{robins1994estimation,bang2005doubly,robins2007comment,cao2009improving,tan2010bounded,van2011targeted,vermeulen2015bias},
among many others.

When either the propensity score model or the outcome model is
misspecified, Propositions~\ref{thmmis}~and~\ref{thmmis2} provide a
complete characterization of the asymptotic behavior of our estimator. In the
same context, \cite{farrell2015robust} proved that the AIPW estimator
is consistent, but Theorem~2 of that work does not yield
an explicit convergence rate. In fact, we show in the supplementary
material that the AIPW estimator has the same convergence rate as
in~\eqref{eqthmmis1}, which is slower than $n^{-1/2}$, and thus the
confidence intervals for the treatment effect are not available under
model misspecification.  In contrast, our estimator is root-$n$
consistent, which leads to honest confidence intervals as shown in
Sections~\ref{secrobust}~and~\ref{secrobust2}. Indeed, this robustness
of the asymptotic distributions to model specification is the main advantage over the AIPW estimators
\citep{belloni2013program,farrell2015robust} and the double selection
estimator \citep{belloni2014inference}.

Unlike our work, the approximate residual balancing method proposed by
\cite{athey2016efficient} does not require the propensity score model
to be sparse or even well formulated. Thus, their method is robust to
the misspecification of the propensity score model under the
assumption that the outcome model is correct. In contrast, our method
requires both models to be sparse. The advantage of our framework is
that it tolerates the misspecified outcome model, so long as the
propensity score model is correctly specified. Thus, our work and
\cite{athey2016efficient} are complementary to one another. In
addition, when the propensity score model is correctly specified and
is indeed sparse, the estimation of the propensity score can help
scientists better understand the treatment assignment mechanism
\citep[e.g.,][]{rubi:08}.  Recall that
$\tilde\pi_i=\pi(\tilde\beta^\top X_i)$. As a byproduct of Theorem
\ref{thm1}, our estimated propensity score is uniformly consistent,
\begin{equation*}
\max_{1\leq i\leq n}|\tilde \pi_i-\pi^*_i|\ = \ O_p\left\{\frac{(s_1\vee s_2)\log (d\vee n)({\log n})^{1/2}}{{n}^{1/2}}\right\}.
\end{equation*}
Thus, the estimated propensity score $\tilde\pi_i$ is an accurate
approximation to the unknown treatment assignment mechanism.  In
contrast, the approximate residual balancing method does not yield an
estimate of the propensity score.  Finally, we note that the linearity
assumption of the outcome model plays an important role in
\cite{athey2016efficient} and their approximate residual balancing
method is not readily applicable if the outcome model is nonlinear.
In contrast, our method is robust to the misspecification of the
outcome model and also can be generalized to nonlinear outcome models.
As an illustration of its generalizability, we consider the extension
of the proposed methodology to the generalized linear models in
Section~\ref{secgeneral}.

In another recent work, \cite{zhao2016covariate} proposed a
generalized covariate balancing method based on a class of scoring
rules. Many existing covariate balancing estimators can be treated as
the primal or dual problems of their optimization
problem. \cite{zhao2016covariate} studied the robustness of these
estimators to misspecified propensity score models under the constant
treatment effect model $\EE\{Y (1)-Y (0)\mid X\} = \tau^*$ for some
constant $\tau^*$. In contrast, our methodology allows for the
heterogeneity of causal effects. In addition, while our work mainly
focuses on the high-dimensional settings, \cite{zhao2016covariate}
does not provide statistical guarantees in such settings.

Most recently, \cite{tan2017regularized,tan2018model} proposed a
penalized calibrated propensity score method and studied its
robustness to model misspecification. Our work is closely related to
\cite{tan2017regularized}, which can be seen as equivalent to directly
plugging the initial estimator $\hat\beta$ into the Horvitz-Thompson
estimator with $w_1(u)=1$.  However, this method does not balance
the covariates as we did in Step 3. Corollary~3 of \cite{tan2017regularized} implies that the
estimator has the rate of the convergence $O_p((s_1\log d/n)^{1/2})$,
which is slower than that of our estimator. In our proof, one can
treat
$\sum_{i=1}^n\left({T_i}/{\hat\pi_i}-1\right)\alpha^{*\top}X_{i}$ as
the ``bias" of the Horvitz-Thompson estimator, which is eliminated by
the covariate balancing step, whereas this term remains in
\cite{tan2017regularized}. In the followup paper, \cite{tan2018model}
removed this bias by constructing an AIPW estimator so that the
resulting estimator is robust to model misspecification. However, our
result is more general than that of \cite{tan2018model}. First, our
Theorem~\ref{thm1}, and Propositions~\ref{thmmis}~and~\ref{thmmis2}
show that there exists a large class of estimators that is
asymptotically normal under possible model misspecification. Second, our theory
holds for generalized linear models as shown in Theorem~\ref{thm2},
whereas \cite{tan2018model}'s method is invalid if the propensity score
model is misspecified.

\section{Covariate Balancing for Generalized Linear
  Models} \label{secgeneral}
\subsection{Method}
In this section, we extend our method to the setting in
which the outcome follows a generalized linear model.  The validity of
many existing methods such as those proposed by \cite{MR3153941},
\cite{fan2016}, and \cite{athey2016efficient} critically rely on the
assumption that the outcome follows a linear model with covariates
$ X _i$ or some transformations (e.g., spline basis) of $ X _i$.
Thus, generalizing the \hCBPS\ to non-linear models is an important
extension.

Assume that the working model for $Y_i(1)$ given $ X _i$ belongs to the exponential family,
\begin{equation}\label{eqglm}
p(y\mid X ) \ = \ h(y,\phi)\exp\left[\frac{1}{a(\phi)}\{y\alpha^{*\top} X -b(\alpha^{*\top} X )\}\right]
\end{equation}
where $h(\cdot,\cdot)$, $a(\cdot)$ and $b(\cdot)$ are known functions,
$\phi$ is the dispersion parameter, and $\alpha^*$ is a
$d$-dimensional vector of unknown regression parameters. For
simplicity, we assume that the dispersion parameter $\phi$ is known.
Given this setup, we propose the following modification of the \hCBPS\
methodology described in Section~\ref{secmethod}.

\begin{description}
\item [Step 1:] Fit the outcome model via the penalized
  maximum likelihood method within the treatment group,
	$$
	\hat\alpha \ = \
        \argmin_{\alpha\in\RR^{d}}\left[-\frac{1}{n}\sum_{i=1}^n
          \frac{T_i}{a(\phi)}\{Y_i\alpha^{\top} X _i-b(\alpha^{\top} X _i)\}+\lambda_0\|\alpha\|_1\right]
	$$
	where $\lambda_0>0$ is a tuning parameter. 
\item [Step 2:] This step is identical to Step~1 in
  Section~\ref{secmethod}, where the weight function $w_1(u)$ is replaced by $w_1(\hat\alpha^{\top}X_i, u)$.  This defines the initial estimator $\hat\beta$.
\item [Step 3:] Re-estimate the outcome model via the penalized weighted 
  maximum likelihood method within the treatment group,
	$$
	\tilde\alpha \ = \
        \argmin_{\alpha\in\RR^{d}}\left[-\frac{1}{n}\sum_{i=1}^n
          \frac{T_i w_2(\hat\beta^\top X_i)}{a(\phi)}\{Y_i\alpha^{\top} X _i-b(\alpha^{\top} X _i)\}+\lambda'\|\alpha\|_1\right]
	$$
	where $\lambda'>0$ is a tuning parameter and $w_2(\cdot)$ is the weight function similar to Step~2 in
  Section~\ref{secmethod}. 
      \item [Step 4:] Define $\tilde S=\{j: |\tilde\alpha_j|>0\}$ and 
        $f(X)=b''(\tilde\alpha^{\top} X ) X_{\tilde
          S}$. Compute,
\begin{equation}\label{eqadaptive}        
\tilde\gamma \ = \ \argmin_{\gamma\in\RR^{|\tilde
    S|}}\|g_n(\gamma)\|^2_2 ~~ {\rm where} ~~
g_n(\gamma)\ = \ \frac{1}{n}\sum_{i=1}^n \left\{\frac{T_i}{\pi(\gamma^\top\bar X _{i\tilde S}+\hat\beta^\top_{\tilde S^c} X _{i\tilde S^c})}-1\right\} f ( X _i).
\end{equation}
Set $\tilde\beta=(\tilde\gamma, \hat\beta_{\tilde S^c})$ and
$\tilde\pi_i=\pi(\tilde\beta^\top X _i)$.
\item [Step 5:] Estimate $\mu_1^*$ by 
  $\hat \mu_1=\frac{1}{n} \sum_{i=1}^{n}{T_iY_i}/{\tilde\pi_i}-\frac{1}{n} \sum_{i=1}^{n}({T_i}/{\tilde\pi_i}-1)b'(\tilde\alpha^{\top} X_i)$.
\end{description}

The current algorithm differs from that in Section~\ref{secmethod} in
the following three ways.  First, Step~1 yields an initial estimator
of $\alpha$, which is then incorporated into the weight function
$w_1(\hat\alpha^{\top}X_i, u)$. As shown later in Theorem~\ref{thm2},
the choice of the weight function $w_1(\hat\alpha^{\top}X_i, u)$
becomes critical when analyzing the asymptotic distribution of
$\hat\mu_1$ under misspecified propensity score
models. 

Second, Step~4 balances the weighted covariates
$f(X)=b''(\tilde\alpha^{\top} X ) X_{\tilde S}$ instead of
$ X _{i \tilde S}$ as done in equation~\eqref{eqcovariatebalance}.
The reason is that to achieve a similar weak covariate balancing
property, one must balance a vector of functions $f( X )$ such that
$b'(\alpha^{*\top} X )\in \textrm{span}\{ f ( X )\}$ where
$\textrm{span}\{ f ( X )\}$ represents the linear space generated by
the basis functions $ f ( X )$. Let $S$ denote the support set for
$\alpha$, i.e., $S=\{j: |\alpha^*_j|>0\}$. Since
$b'(\alpha^{*\top} X )$ is unknown in practice, we approximate
$b'(\alpha^{*\top} X )=b'(\alpha^{*\top}_S X _S)$ by a local linear
estimator
$b'(\tilde\alpha^\top X )+b''(\tilde\alpha^\top X
)(\alpha^*-\tilde\alpha)_S X_S $. Furthermore, if we replace $S$ by an
estimator $\tilde S$, this leads to the weighted covariates
$f(X)=b''(\tilde\alpha^{\top} X ) X_{\tilde S}$ in Step 4.
Unfortunately, balancing $f(X)$ alone does not attain the
(approximate) weak covariate balancing property, because the leading
term $b'(\tilde\alpha^\top X )$ in the local linear approximation has
not been considered. It is possible to add $b'(\tilde\alpha^\top X )$
into the covariate balancing equation, which leads to
$f(X)=\{b'(\tilde\alpha^\top X ), b''(\tilde\alpha^{\top} X )
X_{\tilde S}\}$. However, doing so leads to additional technical
assumptions on the eigenvalues of
$f(X)$. 
To avoid such assumptions, we only attain ``partial" covariate
balancing in Step~4 by choosing $f(X)=b''(\tilde\alpha^{\top} X ) X_{\tilde S}$. 

Third, Step~5 applies the AIPW estimator rather than the
Horvitz-Thompson estimator used in Section~\ref{secmethod}. The
additional term, i.e.,
$\frac{1}{n}
\sum_{i=1}^{n}({T_i}/{\tilde\pi_i}-1)b'(\tilde\alpha^{\top} X_i)$,
comes from the bias due to the imbalance of
$b'(\tilde\alpha^\top X )$. By equation (\ref{eqAIPW}), the AIPW
estimator agrees with the Horvitz-Thompson estimator when the outcome
model is linear. In this case, we have $b''(u)=1$ and $b'(u)=u$ and
balancing $b''(\tilde\alpha^{\top} X )X_{\tilde S}=X_{\tilde S}$ is
sufficient to remove the imbalance effect of
$b'(\tilde\alpha^\top X )=\tilde\alpha^\top X$. Thus, as expected, the
current algorithm reduces to the one in Section~\ref{secmethod} under
the linear outcome model. In addition, if the outcome model is the Poisson regression, we
can also apply the Horvitz-Thompson estimator because under this model
$b''(u)=b'(u)=\exp(u)$ and therefore balancing
$b''(\tilde\alpha^{\top} X )X_{\tilde S}$ is sufficient, provided the intercept term is included.

\subsection{Theoretical Results}


\begin{assumption}[Sub-Exponential condition]\label{asssubx}
  Assume that $\epsilon_1=Y(1)-b'(\alpha^{*\top} X )$
  satisfies $\|\epsilon_1\|_{\psi_1}\leq C$ and $\max_{1\leq i\leq n, 1\leq j\leq d}|X_{ij}|\leq C_n$, where $C$ is a
  positive constant and we allow $C_n$ to increase with $n$.
\end{assumption}	

\begin{assumption}[Sparsity]\label{asssparse2}
Let us denote $s_1=\|\beta^*\|_0$ and $s_2=\|\alpha^*\|_0$. Assume that $C_n^2(s_1\vee s_2)\log (d\vee n)/n^{1/2}=o(1)$, where $C_n$ is defined in Assumption \ref{asssubx}. 
\end{assumption}

\begin{assumption}[Propensity score, outcome model and weight functions]\label{assweight2}
Assume that $Q_n(\beta)$ is a concave function. Let $C,C'$ denote positive constants, which may change from line to line. 
\begin{itemize}
\item[(1)] The same condition (1) in Assumption~\ref{assweight} holds
  for the propensity score model $\pi(u)$.
\item[(2)] In the outcome model, $b(u)$ is third order continuously differentiable and $|X_i^\top\alpha^*|\leq C'$.
\item[(3)] The weight function $w_1(u,v)$ satisfies the following conditions in a small neighborhood of $u^*=X_i^\top\alpha^*$ and $v^*=X_i^\top\beta^*$: $C \leq w_{1}(u,v)\leq 1/C$, $0\leq w'_1(u,v)\leq 1/C$, and the Lipschitz condition in $u$, $|w_1(u,v^*)-w_1(u',v^*)|\leq C'|u-u'|$, where $u\in [u^*-r, u^*+r]$, $v\in [v^*-r, v^*+r]$ for some small constant $r>0$ and $w_1'(u,v)=\partial w_1(u,v)/\partial v$.  
\item[(4)] The same condition (3) in Assumption \ref{assweight} holds for the weight $w_2(u)$.
\end{itemize}
\end{assumption}

Unlike the sub-Gaussian condition in Assumption~\ref{asssubg}, we
allow the error $\epsilon_1$ to be sub-exponential in
Assumption~\ref{asssubx}.  This extension is necessary because many
examples of generalized linear models (e.g., exponential regression
and Poisson regression) satisfy the sub-exponential condition but not
the sub-Gaussian condition. We also allow $C_n$ to possibly grow with
$n$. Specifically, when $X_{ij}$ is uniformly bounded, $C_n$ is a
positive constant. When $X_{ij}$ is sub-Gaussian, then
$\max_{1\leq i\leq n, 1\leq j\leq d}|X_{ij}|=O_p(\{\log
(nd)\}^{1/2})$. Assumption~\ref{asssparse2} requires a similar
sparsity condition, and allows $C_n$ to increase with $n$. Part
(1)~and~(4) of Assumption~\ref{assweight2} are identical to
Assumption~\ref{assweight}. Part~(2) is a mild condition, stating that
the regression effect in the outcome model is bounded. The third order
differentiability of $b(u)$ holds for most generalized linear
models. Part~(3) is a technical condition. To analyze the estimator
$\hat\beta$ in Step~2, we need to control $w_1(u,v)$ and $w'_1(u,v)$
in a small neighborhood of the true values. This condition holds for
two important examples $w_1(u,v)=\pi(v)$ and $w_1(u,v)=b''(u)$. The
former corresponds to example~(a) in Section~\ref{secmethod}, and the
latter represents the generalization of example~(b) to the generalized
linear models.


To study the performance of our estimator under misspecified models,
as done in Sections~\ref{secrobust}~and~\ref{secrobust2}, we define
the least false parameters as,
$$
\beta^o \ = \ \argmax \EE\left[\int_0^{\beta^{\top}X_i }\left\{\frac{T_i}{\pi(u)}-1\right\}w_1(X_i^\top\alpha^*,u) du\right],
$$
$$
\alpha^o \ = \ \argmin \EE\left\{\frac{T_i w_2(\beta^{*\top} X_i)}{a(\phi)}\{Y_i\alpha^{\top} X _i-b(\alpha^{\top} X _i)\}\right\}.
$$
The following main theorem in this section establishes the asymptotic
normality of $\hat\mu_1$ when the outcome variable follows a
generalized linear model. When analyzing the theoretical properties
under model misspecification, we replace $\alpha^*$ and $\beta^*$ in
all assumptions with $\alpha^o$ and $\beta^o$.
\begin{theorem}[Asymptotic Properties under the Generalized Linear Models]\label{thm2}
  Suppose that
  Assumptions~\ref{assun},~\ref{asspropen},~\ref{asseigen},~\ref{asssubx},~\ref{asssparse2},~and~\ref{assweight2}
  hold, and the tuning parameters satisfy
  $\lambda_0\asymp\lambda\asymp\lambda'\asymp \{{\log (d\vee
    n)/n}\}^{1/2}$.
\begin{itemize}
\item [(1)] Assume that both the propensity score model~\eqref{eqps}
  and the outcome model~\eqref{eqglm} are correctly specified. Then
  the estimator $\hat{\mu}_1$ with any weight functions $w_1(u,v)$ and
  $w_2(u)$ satisfies
$$
\hat{\mu}_1-\mu_1^* \ = \ \frac{1}{n}\sum_{i=1}^n\left[\frac{T_i}{\pi^*_i}\{Y_i(1)-b'(\alpha^{*\top}X_i) \}+b'(\alpha^{*\top}X_i) -\mu_1^*\right]+o_p(n^{-1/2}),
$$
and $\hat{\mu}_1$ achieves the same semiparametric efficiency bound.
\item[(2)] Assume that the outcome model~\eqref{eqglm} is correctly
  specified, but the propensity score model~\eqref{eqps} is
  misspecified. If we choose $w_1(u,v)=b''(u)$, then for any $w_2(u)$
  we have
$$
\hat{\mu}_1-\mu_1^* \ = \ \frac{1}{n}\sum_{i=1}^n\left[\frac{T_i}{\pi(X_i^\top\beta^o)}\{Y_i(1)-b'(\alpha^{*\top}X_i) \}+b'(\alpha^{*\top}X_i) -\mu_1^*\right]+o_p(n^{-1/2}).
$$
\item [(3)] Suppose that the propensity score model~\eqref{eqps} is
  correctly specified, but the outcome model~\eqref{eqglm} is
  misspecified. If we set $w_2(u)=\pi'(u)/\pi^2(u)$, then for any
  $w_1(u,v)$ we have
$$
\hat{\mu}_1-\mu_1^* \ = \ \frac{1}{n}\sum_{i=1}^n\left[\frac{T_i}{\pi^*_i}\{Y_i(1)-b'(\alpha^{o\top}X_i) \}+b'(\alpha^{o\top}X_i) -\mu_1^*\right]+o_p(n^{-1/2}).
$$
\end{itemize}
\end{theorem}

Part~(1) of Theorem~\ref{thm2} is the extension of Theorem~\ref{thm1}
to the generalized linear models. Under the correct model
specification, the asymptotic normality of $\hat\mu_1$ holds for any
weight functions $w_1(u,v)$ and $w_2(u)$ that satisfy Assumption
\ref{assweight2}. This result agrees with the theory of AIPW
estimators in \cite{belloni2013program} and
\cite{farrell2015robust}. Unlike the existing work, parts~(2)~and~(3)
provide novel results on the asymptotic normality of $\hat\mu_1$ when
either the propensity score model or the outcome model is
misspecified. Similar to Propositions~\ref{thmmis}~and~\ref{thmmis2},
these results hold only if particular forms of $w_1(u,v)$ and $w_2(u)$
are chosen to remove the bias from model misspecification. In particular, we use
the weight $w_1(u,v)=b''(u)$ in part~(2), which requires the knowledge
of $\alpha^*$ in the outcome model. This explains why Step~1 is
needed. Since part~(2) holds for any weight function $w_2(u)$, the
estimator $\hat{\mu}_1$ remains asymptotically normal even if we skip
Step~3 and replace $\tilde\alpha$ in Step~4 with $\hat\alpha$ in
Step~1. Similarly, part~(3) holds for any weight function
$w_1(u,v)$. Thus, if we set $w_2(u)=\pi'(u)/\pi^2(u)$ and
$w_1(u,v)=\pi(v)$, we may skip Step~1 of our algorithm and 
the same result in part~(3) still applies.

Similar to Remark~\ref{remrobust}, the proposed estimator $\hat\mu_1$
when $w_1(u,v)=b''(u)$ and $w_2(u)=\pi'(u)/\pi^2(u)$ is asymptotically
normal provided that either the propensity score model or the outcome
model is correctly specified. This estimator does not require to know
which of the two models is correct, and therefore is recommended for
 practical use.

\section{Simulation Studies}\label{secnum}
In this section, we conduct simulation studies to evaluate the finite
sample performance of the proposed \hCBPS\ methodology. We consider
the following data generating processes. First, we generate the $d$
dimensional covariate $X_i\sim N(0,\Sigma)$ where
$\Sigma_{jk}=\rho^{|j-k|}$ with $\rho=1/2$.  We generate the binary
treatment $T_i$ using the logistic regression model of the form,
$\pi(X_i) = 1-1 / \{1 + \exp(-X_{i1}
+X_{i2}/2-X_{i3}/4-X_{i4}/10-X_{i5}/10+X_{i6}/10)\}$. For the
potential outcomes, we consider both linear and logistic regression
models as specified later. The observed outcome is
$Y_i=Y_i(1)T_i+Y_i(0)(1-T_i)$.

The simulation is repeated 200 times under each setting. Throughout
the simulation studies, whenever possible, we compare our method
(\hCBPS) to the approximate residual balancing (\RB) method
\citep{athey2016efficient}, the regularized AIPW (\AIPW) method
\citep{farrell2015robust,belloni2013program} and the double selection
\citep{belloni2014inference}. For the sake of comparison, we use the
Lasso penalty in both \hCBPS\ and \AIPW\ methods, and all tuning
parameters are determined by the 5 fold cross-validation. The weight
functions in our method are chosen according to Remark
\ref{remrobust}. For the \RB\ method, we use the default values of the
tuning parameters in the \R\ package {\sf balanceHD}. The double
selection method is implemented using the \R\ package {\sf hdm}.


We first consider the setting, in which the potential outcomes are
generated from the linear regression models:
\begin{equation*}
\begin{split}
  Y_i(1)  \ = \ 2 &+ 0\cdot 137(X_{i5} + X_{i6}+ X_{i7}+X_{i8})+ \epsilon_{1i},\\
  Y_i(0)  \ = \ 1 &+ 0\cdot 291(X_{i5} + X_{i6} + X_{i7} +X_{i8}+ X_{i9}+X_{i10})+ \epsilon_{0i},\\
\end{split}
\end{equation*}
where $\epsilon_{1i}$ and $\epsilon_{0i}$ are independent standard
normal random variables.  Under this setting, we consider the
following four scenarios.  In the first scenario, we assume that the
propensity score and outcome models are correctly specified.  In the
second scenario, the outcome models are correctly specified but the
propensity score model is misspecified. We use the transformed
variables,
$X_{mis} = \{\exp(X_1/2), X_2/\{1+\exp(X_1)\}+10,
(X_1X_3/25+0\cdot6)^3, (X_2+X_4+20)^2, X_6, \exp(X_6+X_7), X_9^2,
X_7^3-20, X_9, \cdots X_d\}$ to generate the treatment but the
original variables $X$ to generate the outcome variables.  In the
third scenario, the propensity score model is correctly specified but
the outcome models are misspecified. We use the same transformed
variables $X_{mis}$ to generate the outcomes but the original
variables $X$ to generate the treatment. Finally, we consider a
scenario, in which both the outcome and propensity score models are
misspecified using the transformed covariates. This model
misspecification follows the work of \cite{kang2007demystifying} who
evaluated the empirical performance of the AIPW estimator in
low-dimensional settings.

\begin{table}[t!]
\singlespacing
\small
\caption{Bias, standard error (Std Err), standardized root-mean-squared error (RMSE), coverage probability of 95\% confidence intervals (Coverage), and length of 95\% confidence intervals (CI length) for the estimation of
  the ATE.  Four methods --
  high-dimensional CBPS, approximate residual
  balancing, regularized augmented inverse
  probability weighting, and double selection  -- are compared.  
 }\label{tab1}\vspace{-.1in}
  \begin{center}
  \scalebox{0.8}{
    \begin{tabular}{rcccccccccc}
      \toprule 
$n=500$    &\multicolumn{4}{c}{$d=1000$} &\multicolumn{4}{c}{$d=2000$} \\		
      \cmidrule(r){2-5}\cmidrule(r){6-9}		
         &  \hCBPS &  \RB    &  \AIPW  & \DS   &  \hCBPS &  \RB    &  \AIPW &  \DS \\
      \midrule
      \multicolumn{5}{l}{\it (1) Both models are correct} \\
      Bias &   -0$\cdot$0026 & -0$\cdot$0017 & -0$\cdot$0498 & -0$\cdot$0910       &-0$\cdot$0595 & -0$\cdot$0580 & -0$\cdot$1200 & {-0$\cdot$0397} \\ 
      Std Err &  0$\cdot$0936 & 0$\cdot$1074 & {0$\cdot$0926} & 0$\cdot$0979       & 0$\cdot$1061 & 0$\cdot$1155 & {0$\cdot$1011} & 0$\cdot$1279\\
      R{MSE} &  {0$\cdot$0936} & 0$\cdot$1074 & 0$\cdot$1052 & 0$\cdot$1337    &  0$\cdot$1216 & 0$\cdot$1292 & 0$\cdot$1569 & 0$\cdot$1334  \\
      Coverage & 0$\cdot$965 & 0$\cdot$930 &0$\cdot$915 & 0$\cdot$890              & 0$\cdot$910 & 0$\cdot$910 &0$\cdot$855 & 0$\cdot$945 \\
      CI length &  0$\cdot$3867 & 0$\cdot$4231 & 0$\cdot$3775 & 0$\cdot$4294           &  0$\cdot$3862 & 0$\cdot$4359 & 0$\cdot$3731 & 0$\cdot$5034   \\
       \midrule     
      \multicolumn{5}{l}{\it (2) Propensity score model is misspecified} \\
      Bias &   {-0$\cdot$0120} & -0$\cdot$0303 & -0$\cdot$1078 & -0$\cdot$0782     &-0$\cdot$0446& -0$\cdot$0685& -0$\cdot$1234 &   {-0$\cdot$0357} \\ 
      Std Err &  0$\cdot$0984 & 0$\cdot$1153 & {0$\cdot$0963} & 0$\cdot$1034   &0$\cdot$0924& 0$\cdot$1041& {0$\cdot$0921} & 0$\cdot$1214 \\
      R{MSE} &  {0$\cdot$0991} & 0$\cdot$1193 & 0$\cdot$1446 & 0$\cdot$1296   &{0$\cdot$1025}& 0$\cdot$1246& 0$\cdot$1540& 0$\cdot$1265 \\
      Coverage & 0$\cdot$965 & 0$\cdot$945 &0$\cdot$815 & 0$\cdot$905 &0$\cdot$930& 0$\cdot$910& 0$\cdot$740& 0$\cdot$940 \\
      CI length &  0$\cdot$3864 & 0$\cdot$4431 & 0$\cdot$3732 & 0$\cdot$4227   & 0$\cdot$3839& 0$\cdot$4382 & 0$\cdot$3702& 0$\cdot$5023\\
      \midrule
      \multicolumn{5}{l}{\it (3) Outcome model is misspecified} \\
      Bias &   {-0$\cdot$0034} & -0$\cdot$0321 & -0$\cdot$0562 & -0$\cdot$0991   &{-0$\cdot$0317}& -0$\cdot$0572& -0$\cdot$1215& -0$\cdot$0443  \\ 
      Std Err &  0$\cdot$0917 & 0$\cdot$0982 & {0$\cdot$0914} & 0$\cdot$1023     &0$\cdot$0944& 0$\cdot$0992& {0$\cdot$0921}& 0$\cdot$1026 \\
      R{MSE} &  {0$\cdot$0917} & 0$\cdot$1033 & 0$\cdot$1072 & 0$\cdot$1424   &{0$\cdot$0995}   & 0$\cdot$1145& 0$\cdot$1525& 0$\cdot$1118\\
      Coverage & 0$\cdot$960 & 0$\cdot$960 &0$\cdot$905 & 0$\cdot$845       &0$\cdot$950   & 0$\cdot$955& 0$\cdot$770& 0$\cdot$945 \\
      CI length &  0$\cdot$3874 & 0$\cdot$4292 & 0$\cdot$3815 & 0$\cdot$4327  &0$\cdot$3890& 0$\cdot$4403& 0$\cdot$3728& 0$\cdot$4261\\    
      \midrule
      \multicolumn{5}{l}{\it (4) Both models are misspecified} \\
      Bias &   {-0$\cdot$0547} & -0$\cdot$1201 & -0$\cdot$1873 & -0$\cdot$1005    & {-0$\cdot$0243} & -0$\cdot$0599& -0$\cdot$1393& -0$\cdot$0518\\ 
      Std Err &  0$\cdot$1106 & 0$\cdot$1038 & {0$\cdot$0903} & 0$\cdot$0950  &0$\cdot$0969& 0$\cdot$1060& {0$\cdot$0921} & 0$\cdot$0965 \\
      R{MSE} &  {0$\cdot$1234} & 0$\cdot$1588 & 0$\cdot$2079 & 0$\cdot$1383 &{0$\cdot$0999}& 0$\cdot$1218& 0$\cdot$1670& 0$\cdot$1095\\
      Coverage & 0$\cdot$890 & 0$\cdot$815 &0$\cdot$775 & 0$\cdot$875  &0$\cdot$940& 0$\cdot$940& 0$\cdot$720& 0$\cdot$950\\
      CI length &  0$\cdot$3994 & 0$\cdot$4586 & 0$\cdot$3790 & 0$\cdot$4333   &0$\cdot$3948& 0$\cdot$4545& 0$\cdot$3781& 0$\cdot$4334\\
\midrule

$n=1000$    &\multicolumn{4}{c}{$d=1000$} &\multicolumn{4}{c}{$d=2000$} \\		
      \cmidrule(r){2-5}\cmidrule(r){6-9}		
         &  \hCBPS &  \RB    &  \AIPW  & \DS   &  \hCBPS &  \RB    &  \AIPW &  \DS \\
      \midrule     
      \multicolumn{5}{l}{\it (1) Both models are correct} \\
      Bias &   {-0$\cdot$0233} & -0$\cdot$0234 & -0$\cdot$0814 & -0$\cdot$0476       &0$\cdot$0199 & 0$\cdot$0186 & {-0$\cdot$0056} & 0$\cdot$0249 \\ 
      Std Err &  0$\cdot$0669 & 0$\cdot$0777 & {0$\cdot$0647} & 0$\cdot$0690        &0$\cdot$0659 & 0$\cdot$07476& {0$\cdot$0654}& 0$\cdot$0757 \\
      {RMSE} &  0$\cdot$0695 & 0$\cdot$0729 & 0$\cdot$0678 & 0$\cdot$0839        &0$\cdot$0689& 0$\cdot$0769& {0$\cdot$0657} & 0$\cdot$0797 \\
      Coverage & 0$\cdot$955 & 0$\cdot$940 &0$\cdot$905 & 0$\cdot$920                  &0$\cdot$940& 0$\cdot$935& 0$\cdot$950& 0$\cdot$955 \\
      CI length &  0$\cdot$2828 & 0$\cdot$3010 & 0$\cdot$2700 & 0$\cdot$2978             &0$\cdot$2746 & 0$\cdot$2979& 0$\cdot$2697& 0$\cdot$3187\\
      \midrule
      \multicolumn{5}{l}{\it (2) Propensity score model is misspecified} \\
      Bias &   {-0$\cdot$0297} & -0$\cdot$0455 & -0$\cdot$0931 & -0$\cdot$0362     &0$\cdot$0164& {0$\cdot$0135}& -0$\cdot$0137 &   0$\cdot$0116 \\ 
      Std Err &  0$\cdot$0607 & 0$\cdot$0694 & {0$\cdot$0605} & 0$\cdot$0665           &0$\cdot$0662& 0$\cdot$0758& {0$\cdot$0659} & 0$\cdot$0846 \\
      R{MSE} &  {0$\cdot$0671} & 0$\cdot$0842 & 0$\cdot$1105 & 0$\cdot$0757     &0$\cdot$0682& 0$\cdot$0770& {0$\cdot$0673}& 0$\cdot$0854 \\
      Coverage & 0$\cdot$970 & 0$\cdot$930 &0$\cdot$855 & 0$\cdot$930                    &0$\cdot$940& 0$\cdot$955& 0$\cdot$955& 0$\cdot$935 \\
      CI length &  0$\cdot$2801 & 0$\cdot$3040 & 0$\cdot$2694 & 0$\cdot$2960            & 0$\cdot$2746& 0$\cdot$2987 & 0$\cdot$2697& 0$\cdot$3381\\
      \midrule
      \multicolumn{5}{l}{\it (3) Outcome model is misspecified} \\
      Bias &   {-0$\cdot$0222} & -0$\cdot$0229 & -0$\cdot$0821 & -0$\cdot$0436          &{-0$\cdot$0062}& -0$\cdot$0026& -0$\cdot$0517& 0$\cdot$0262  \\ 
      Std Err &  0$\cdot$0670 & 0$\cdot$0699 & {0$\cdot$0653} & 0$\cdot$0671      &0$\cdot$0653& 0$\cdot$0709& {0$\cdot$0630}& 0$\cdot$0669 \\
      R{MSE} &  {0$\cdot$0706} & 0$\cdot$0735 & 0$\cdot$1049 & 0$\cdot$0800     &{0$\cdot$0656}   & 0$\cdot$0709& 0$\cdot$0815& 0$\cdot$0718\\
      Coverage & 0$\cdot$960 & 0$\cdot$960 &0$\cdot$890 & 0$\cdot$955               &0$\cdot$975    & 0$\cdot$970& 0$\cdot$930& 0$\cdot$965  \\
      CI length &  0$\cdot$2842 & 0$\cdot$3058 & 0$\cdot$2709 & 0$\cdot$3002               &0$\cdot$2848& 0$\cdot$3139& 0$\cdot$2726& 0$\cdot$2920 \\    
      \midrule
      \multicolumn{5}{l}{\it (4) Both models are misspecified} \\
      Bias &   -0$\cdot$0157 & {-0$\cdot$0072} & -0$\cdot$0504 & -0$\cdot$0366       & 0$\cdot$0150 & {0$\cdot$0009}& -0$\cdot$0635& 0$\cdot$0076 \\ 
      Std Err &  {0$\cdot$0701} & 0$\cdot$0822 & 0$\cdot$0721 & 0$\cdot$0687           &0$\cdot$0613& 0$\cdot$0765& {0$\cdot$0598}& 0$\cdot$0792 \\
      R{MSE} &  {0$\cdot$0718} & 0$\cdot$0825 & 0$\cdot$0880 & 0$\cdot$0779    &{0$\cdot$0631}& 0$\cdot$0765& 0$\cdot$0872 & 0$\cdot$0796\\
      Coverage & 0$\cdot$945& 0$\cdot$960 &0$\cdot$905 & 0$\cdot$925                        &0$\cdot$990& 0$\cdot$960& 0$\cdot$905& 0$\cdot$950\\
      CI length &  0$\cdot$2872 & 0$\cdot$3117 & 0$\cdot$2774 & 0$\cdot$3046              &0$\cdot$2882& 0$\cdot$3281& 0$\cdot$2739& 0$\cdot$3426\\

      \bottomrule
    \end{tabular}
    }
  \end{center}
\vspace{-.2in}

\end{table}

Table~\ref{tab1} shows the bias, standard error, standardized root
mean squared error (RMSE)
$\{{\mathbb{E}(\hat{\mu} - \mu)^2}\}^{1/2}/\mu$, coverage probability
of 95\% confidence intervals, and their length 
for the estimation of the average treatment effect under the four
scenarios. We focus on the comparison in the high-dimensional setting
with $d=1000, 2000$ and sample size $n=500, 1000$. Some additional
simulation studies for smaller $d$ is deferred to the supplementary
materials. Table~\ref{tab1} shows that the proposed method tends to have 
 smaller RMSE in most scenarios. More importantly, as seen
in scenarios (2) and (3), the fact that the \hCBPS\ has an accurate
coverage probability under model misspecification provides empirical
support for the robustness property established in
Propositions~\ref{thmmis} and \ref{thmmis2}. In contrast, the AIPW
estimator has a
significant bias under scenarios (2) and (3). As a result, its
coverage probability is below 0$\cdot$95 in most cases.  The other two
methods, i.e., RB and double selection, perform reasonably well under
model misspecification. But their confidence intervals tend to be
wider than the proposed method.


We also consider the simulation with logistic outcome models. When the
outcome variable is binary, RB is not directly applicable. Thus, we
only compare our method with the regularized AIPW and the
double selection method. The simulation results illustrate the same
conclusion. Due to the space constraint, we defer the details to the
supplementary materials.  The
supplementary materials also contain more extensive numerical
results including the simulations under different data generating
processes, non-sparse models,  comparisons with many other
estimators (e.g., normalized Horvitz-Thompson estimator, calibrated
likelihood \citep{tan2010bounded}, targeted maximum likelihood
estimator \citep{van2011targeted}, IPW, and standard CBPS estimators) under a moderate dimension, and sensitivity analysis with respect to the choice of tuning parameters.


In summary, the proposed \hCBPS\ estimator tends to have a smaller
mean squared error, is more robust to model misspecification, and
exhibits accurate coverage probability in finite samples. Our results
are consistent with the empirical findings of \cite{MR3153941} and
\cite{fan2016} that covariate balancing tends to outperform the AIPW
estimator in low-dimensional settings.  Our simulation studies imply
that the same conclusion appears to hold in high-dimensional settings.

\section{Empirical Illustration}\label{secdata}

For empirical illustration, we consider a dataset obtained from the
first two waves of Jennings' and Niemi's Political Socialization Panel
Study, which is originally analyzed by
\cite{kam2008reconsidering}. One purpose of this study is to
understand the effect of higher education on political
participation. The dataset consists of $1,051$ randomly selected high
school seniors in the class of 1965. The information about each sample
is collected via in-person interviews in the first wave of the study,
which we treat as pre-treatment covariates. The second wave of the
study conducted in 1973 collects the outcome variable, political
participation, as well as the dichotomous treatment variable, college
attendance.

For the purpose of comparison, we follow the original study and use 81
pre-treatment covariates, which include gender, race, club
participation, and academic performance. Since many of the covariates
are categorical variables with more than two levels, we create an
indicator variable that represents each level. Therefore, a total of
204 pre-treatment variables are used in the propensity score and
outcome models. The outcome variable represents an index of adult
political participation, which is equal to the sum of eight acts
including the turnout in the 1972 presidential election, attending
campaign rallies, making a donation to a campaign, and displaying a
campaign button and bumper sticker.  Since this variable takes an
integer value ranging from zero to eight, we use the binomial logistic
regression for the outcome model.  The propensity score model is
assumed to be the logistic regression.  We then estimate both the ATE
and ATT of college attendace on political participation (the number of
treated observations is 675).

We apply five methods to analyze this dataset, the proposed \hCBPS\
methodology, the regularized \AIPW\ method \citep{farrell2015robust},
the original \CBPS\ method \citep{MR3153941}, the AIPW method without
regularization (\AIPWNR) \citep{robins1994estimation} and the IPW
estimator with the regularized logistic regression (IPW). The
estimation procedures for the first two methods are identical to those
described in the simulation studies. For the original \CBPS\
methodology, it is designed for the linear outcome model, which does
not provide an ideal balance of pre-treatment variables. In addition,
we use the bootstrap method to approximate the standard error of the
estimator based on the CBPS and IPW methods.


The results are shown in Table~\ref{tab7}. The 
\hCBPS, \AIPW, IPW and \CBPS\ methods imply that the overall ATE of college
education on political participation is positive and statistically
significant while \AIPWNR\ yields a smaller estimate with a larger
standard error.  The ATE estimates and their associated standard
errors based on the regularized methods (i.e., \hCBPS, IPW and \AIPW) are
quite similar to each other.  These ATE estimates are, however,
smaller than that of \CBPS\ and greater than that of \AIPWNR.  More
importantly, for both the ATE and the ATT, the regularized estimate
\hCBPS\ has much smaller standard errors than \CBPS\ and \AIPWNR.
There are at least two reasons for this difference in standard errors.
First, as shown in Section~\ref{secgeneral}, the \hCBPS\ methodology
uses a different covariate balancing estimating equation than \CBPS\
when the outcome model is nonlinear, achieving the semiparametric
efficiency bound. Second, the original \CBPS\ methodology tends to be
unstable when balancing a large number of covariates (204 in this
case). Thus, the proposed \hCBPS\ method improves the existing
covariate balancing methods when the outcome model belongs to the
class of generalized linear models and the number of covariates is
large.

We also apply the \hCBPS, \AIPW, IPW and \AIPWNR\ methods to the subsample
of whites separately.  Among a total of 1,051 respondents, there are
966 white respondents. In this case, the \CBPS\ method does not
converge, and therefore the estimate is unavailable. The results
appear at the last row of Table~\ref{tab7}.  Again, the estimates of
the  two regularized methods are similar so are the standard
errors.  
However, the \AIPWNR\ methodology shows very large
variance, mainly because the maximum likelihood estimate of propensity
score tends to be very unstable when the number of covariates is large.

\begin{table}
\singlespacing
\small
	\caption{The estimated average effects of college attendance on political participation.  The estimates based on the proposed \hCBPS\ methodology are compared with those of the original \CBPS\ estimator, the regularized augmented inverse probability weighted estimator (\AIPW), the augmented inverse probability weighted estimator without regularization (\AIPWNR) and the inverse propensity score weighted estimator with the regularized logistic regression (IPW).  Standard errors appear in parentheses. }
 \label{tab7}\vspace{-.1in}
	\begin{center}
		\begin{tabular}{lccccc}
			\toprule 
& \hCBPS &  \CBPS &  \AIPW  & \AIPWNR & IPW\\
\midrule
\multirow{2}{*}{Overall (ATE)} & 0$\cdot$8293  &1$\cdot$0163 &0$\cdot$8796& 0$\cdot$4904&1$\cdot$0666\\
& (0$\cdot$1247)  &(0$\cdot$2380) &(0$\cdot$1043)& (0$\cdot$6009)&(0$\cdot$1588)\\
\multirow{2}{*}{Overall (ATT)} & 0$\cdot$8439  &1$\cdot$1232 & \\
& (0$\cdot$1420)  & (0$\cdot$3094)&\\
\multirow{2}{*}{Whites (ATE)} & 0$\cdot$8445   & & 0$\cdot$8977&0$\cdot$1205 &1.1371\\
& (0$\cdot$1279)   & & (0$\cdot$1089)& (9$\cdot$4522)& (0$\cdot$1548)\\
 \bottomrule
		\end{tabular}
	\end{center}
        \vspace{-.2in}

\end{table}

\section{Discussion}

There are several future directions that are worthy of further
investigation.  First, it is important to extend these
high-dimensional causal inference methods to non-binary treatment
regimes, including continuous treatment and dynamic treatment regimes.
Second, we plan to further study the effect of tuning parameters on
statistical inference. In numerical experiments, the tuning parameters
are chosen by the cross-validation, which leads to reasonable finite
sample results. Based on the sensitivity analysis, the results appear
to be stable with respect to a small perturbation of tuning
parameters. One interesting problem is to formally justify the
validity of the inference based on the cross-validated estimators. The
current research on the cross-validated Lasso estimator only
guarantees a slow rate of convergence when the error is sub-Gaussian
(e.g., \cite{chetverikov2016cross} and the references therein), which
is not sufficient to control the remainder terms in the proof of
Theorem \ref{thm1}. Further theoretical development is needed to
address this important problem.



\section*{Acknowledgement}
We thank Jamie Robins for the insightful discussion.

\section*{Supplementary material}
Supplementary material includes proofs of the theoretical results, additional simulation results and further technical details.

\bibliographystyle{ims}
\bibliography{cbps,my,imai,spglm,spglm2}

\end{document}